\documentclass[preprint,showpacs,showkeys,preprintnumbers,amsmath,amssymb]{revtex4}

\usepackage[latin1]{inputenc}
\usepackage[active]{srcltx}
\usepackage{setspace}
\usepackage{graphicx}
\usepackage{dcolumn}
\usepackage{bm}
\usepackage{epsfig}
\usepackage{epstopdf}

\begin{document}

\preprint{Preprint IST/DF 1.2014-M J Pinheiro}


\title[]{The Flyby Anomaly and the Effect of a Topological Torsion Current}

\author{Mario J. Pinheiro}
\email{mpinheiro@tecnico.ulisboa.pt}

\affiliation{Departament of Physics, Instituto Superior T\'{e}cnico - IST, Universidade de Lisboa - UL,
Av. Rovisco Pais, 1049-001 Lisboa, Portugal\\
Phone: 351.1.21.841.93.22\\
351.1.21.846.44.55}

\homepage{http://web.ist.utl.pt/d2493/}

\thanks{}

\date{\today}

\begin{abstract}
A new variational technique determines the general condition of equilibrium of a rotating gravito-electromagnetic system and provides a modified dynamical equation of motion from where it emerges a so-far unforseen topological torsion current (TTC) [Mario J. Pinheiro (2013) 'A Variational Method in Out-of-Equilibrium Physical Systems', Scientific Reports {\bf 3}, Article number: 3454]. We suggest that the TTC may explain, in a simple and direct way, the anomalous acceleration detected in spacecrafts during close planetary flybys. In addition, we theorize that TTC may represent an unforeseen relationship between linear momentum and angular motion through the agency of a vector potential.
\end{abstract}



\maketitle

Despite an apparent resolution of the so-called flyby anomaly, we believe that there is still a reason for an ongoing debate about the causes of the anomalous, small and constant Doppler frequency drift shown by the radio-metric data from Pioneer 10$/$11, which can be interpreted as a uniform acceleration of $a_P=(8.74 \pm 1.33) 10^{-8}$ cm$/$s$^2$ towards the Sun~\cite{Turyshev 2007,Nieto 2008,Anderson_2008}, in particular, when it became clear that a number of satellites in Earth flyby have undergone mysterious energy changes~\cite{Anderson_2008}. This effect is essentially a slight departure from Newtonian acceleration and according to Brownstein and Moffat~\cite{Moffat} this acceleration is directed toward to the Sun, according to $a_P=-\delta G(r)M_{\oplus}/r^2$, with $\delta G(r)=G_0 \alpha(r) [1-e^{-r/\lambda(r)} (1+\frac{r}{\lambda(r)})]$.

There is several proposed explanations for this effect, among them we may refer: an adiabatic acceleration of light due to an adiabatic decreasing of the permeability and permittivity of empty space~\cite{Ranada 2003}; the dilaton-like Jordan-Brans-Dicke scalar field as the source of dark energy and giving rise to a new term of force with magnitude $a_P=F_r/m=-c^2/R_H$ ($R_H$ is the Hubble scale), see Ref.~\cite{Castro 2009}; light speed anisotropy~\cite{Cahill 2008} based on Lorentz space-time interpretation and resorting from the earlier measurement od D. C. Miller (see also Ref.~\cite{Rodrigo} which gives an interesting reformulation of special theory of relativity); a computer modeling technique called Phong reflection model~\cite{Bertolami_2010} may apparently explains the effect as mainly due to the heat reflected from the main compartment, but it still needs confirmation.

In this paper, we suggest a possible theoretical explanation of the physical process underlying the unexpected orbital-energy change observed during the close planetary flybys~\cite{Anderson 2008,Turyshev 2009} based on the topological torsion current (TTC) found in a previous work~\cite{Pinheiro_SR}. Anderson et al. ~\cite{Anderson_2003} proposed an helicity-rotation coupling that is more akin to our proposal. However, the anomalous acceleration cannot be explained by means of their mechanism due to its small magnitude.

A standard technique for treating thermodynamical systems on the basis of information-theoretic framework has been developed
previously~\cite{Jaynes,Pinheiro:02,Pinheiro:04,Pinheiro_SR}. We can find in technical literature several textbooks that give an overview over the subject, see e.g., Ref.~\cite{Landau_1,Chandrasekhar1,Khinchin,Liboff,Prigogine:71,Greenspan}. The referred work may be applied to a self-gravitating plasma system, and the extended mathematical formalism developed to investigate out-of-equilibrium systems in the framework of the information theory, can be applied for the analysis of the equilibrium and stability of a gravito-electromagnetic system (e.g., rotating plasma, or spacecraft in a gravitationally-assisted manoeuver).

Our method is fundamentally based on the method of Lagrange multipliers applied to the total entropy of an ensemble of particles. However, we use the fundamental equation of thermodynamics $dU= T d S- \sum_k F_k dx^k$ on differential forms, considering $U$ and $S$ as 0-forms. As we have shown in a previous work~\cite{Pinheiro:04} we obtain a set of two first order differential equations that reveal the same formal symplectic structure shared by classical mechanics, fluid mechanics and thermodynamics.

Following the mathematical procedure proposed in
Ref.~\cite{Pinheiro:04} the total entropy of the system
$\overline{S}$, considered as a formal entity describing an out-of-equilibrium physical system, is given by
\begin{widetext}
\begin{equation}\label{eq2}
\overline{S} = \sum_{\alpha=1}^N \{ S^{(\alpha)} [ E^{(\alpha)} -
\frac{(p^{(\alpha)})^2}{2 m^{(\alpha)}} - q^{(\alpha)} V^{(\alpha)}
+ q^{(\alpha)} (\mathbf{A}^{(\alpha)} \cdot \mathbf{v}^{(\alpha)}) -
m^{(\alpha)} \phi^{(\alpha)} (\mathbf{r}) - m^{(\alpha)}
\sum_{\beta=1}^N \phi^{(\alpha,\beta)}] + (\mathbf{a} \cdot
\mathbf{p}^{(\alpha)} + \mathbf{b} \cdot ([\mathbf{r}^{(\alpha)}
\times \mathbf{p}^{(\alpha)}] ) \}.
\end{equation}
\end{widetext}
Although it has been argued that $S$ was defined for equilibrium states and had no time dependence of any kind, one might think that it must be possible to describe entropy by some means during the evolution of a physical system. But if the time evolution of others physical quantities can be made, like energy $E$, pressure $P$ and number of particles $N$, then why not $S$. As in our previous work~\cite{Pinheiro_SR}, regardless of these uncertainties, the explanation proposed here provides a different input to move further toward a better understanding of the role of entropy.

\begin{figure}
  \includegraphics[width=3.75 in, height=3.7 in]{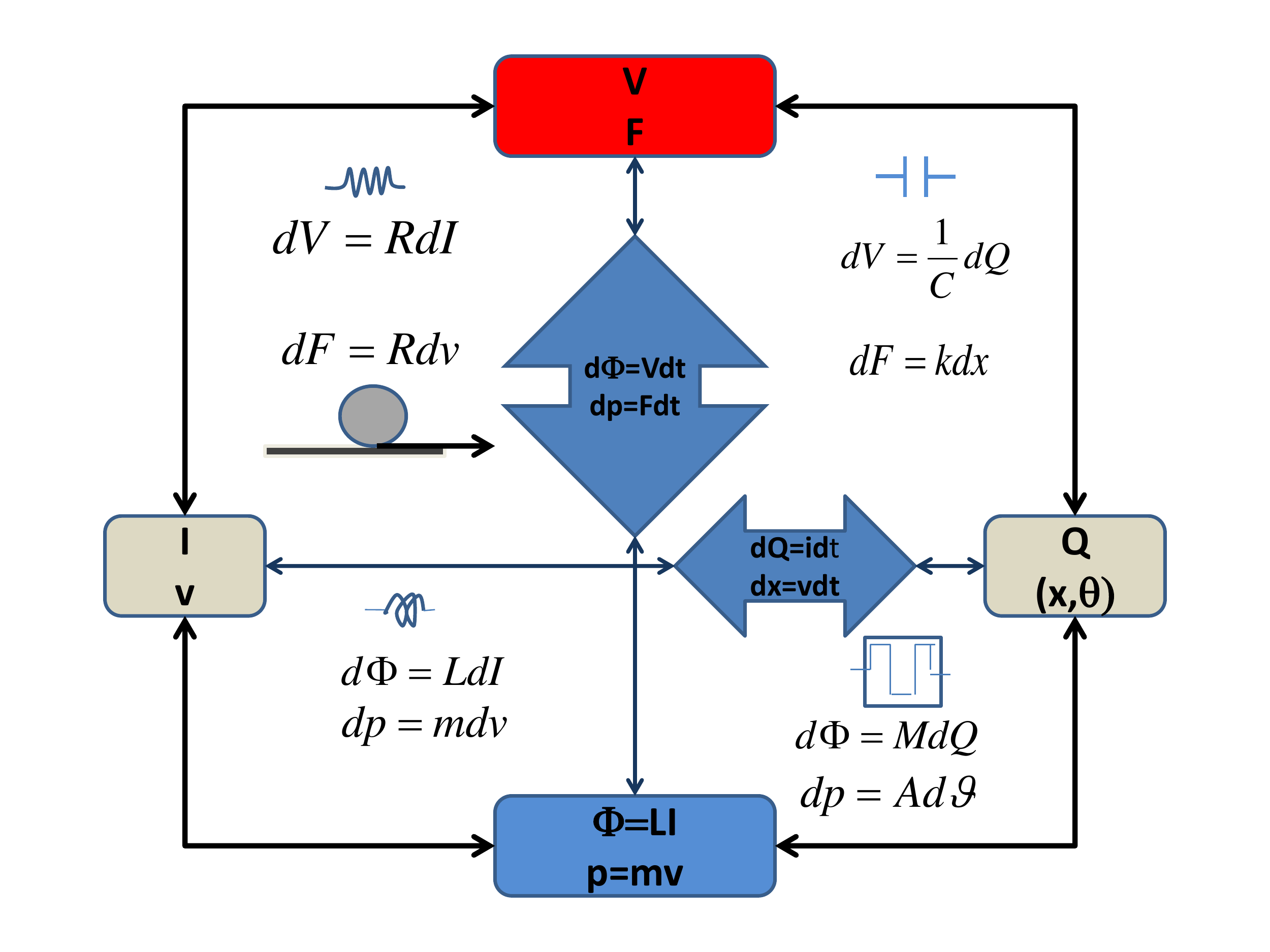}\\
  \caption{The missing fourth element of force: following an analogy with the electromagnetic field, it is expected a new element of force, the topological torsion current. It is shown the standard symbols used for resistors, capacitors, solenoids and memristors.}\label{Fig1}
\end{figure}

The conditional extremum points give the canonical momentum and the dynamical equations of motion of a general physical system in out-of-equilibrium conditions. Then the two first order differential equations can be represented in the form (see Ref.~\cite{Pinheiro:04}):
\begin{align}
\partial_{\mathbf{p}^{(\alpha)}} \bar{S} \geq 0 \label{g2b}\\
\nonumber \\
\partial_{\mathbf{r}^{(\alpha)}} \bar{S} = -\eta \partial_{\mathbf{r}^{(\alpha)}} U^{(\alpha)} - \eta m^{(\alpha)} \partial_t \mathbf{v}^{(\alpha)} \geq 0  \label{g2a}.
\end{align}
Here, $\eta \equiv 1/T$ is the inverse of the "temperature" (not being used so far), and we use condensed notation: $\partial_{\mathbf{p}^{(\alpha)}} \equiv \partial/\ \partial_{\mathbf{p}^{(\alpha)}}$. Then we obtain a general equation of dynamics for electromagnetic-gravitational systems:
\begin{widetext}
\begin{equation}\label{eq7b}
\rho \frac{d \mathbf{v}}{d t} = \rho \mathbf{E} + [\mathbf{J} \times \mathbf{B}] - \pmb{\nabla} \phi -
\pmb{\nabla} p + \rho [\mathbf{A}  \times \pmb{\omega}].
\end{equation}
\end{widetext}
The last term of Eq.~\ref{eq7b} represents the topological torsion current~\cite{Pinheiro:04} (TTC) and we may stress how $\mathbf{A}$ may be considered physically real, even in the frame of the gravitational field, despite eventually the arbitrariness in its divergence. This force does work in order to increase the rotational energy of the system, producing a rocket-like rotation effect on a plasma, or the orbital-energy change observed during the close planetary flybys,
an issue thoroughly discussed in Ref.~\cite{Anderson_2007}. Moreover, the topological torsion current emerge from the universal competition between entropy and energy, each one seeking a different equilibrium condition (this happens in the case of planetary atmospheres, when energy tends to assemble all atmospheric molecules on the surface of the planet, but entropy seeks to spread them evenly in all available space).
This TTC may be envisaged as the missing force term in the traditional hierarchy of agencies responsible for the motion of matter, as depicted in Fig.~\ref{Fig1}, and following along the same electromagnetic analogy proposed by Chua~\cite{Chua_1971}. The basic four physical quantities are the electric current $i$ (or speed $v$), the voltage $V$ (or the force $F$), the charge $q$ (or the position $x$), and the flux-linkage $\Phi$ (or momentum $p=mv$). Under the logical point of view, from six possible combinations among these four variables, five are already well-known. However, the TTC points to the existence of a so-far unforeseen relationship between momentum and angular motion through the agency of a vector potential (see Refs.~\cite{Chua_1971,Abbott_2012}).

\begin{figure}
  \includegraphics[width=3.5 in, height= 3.5 in]{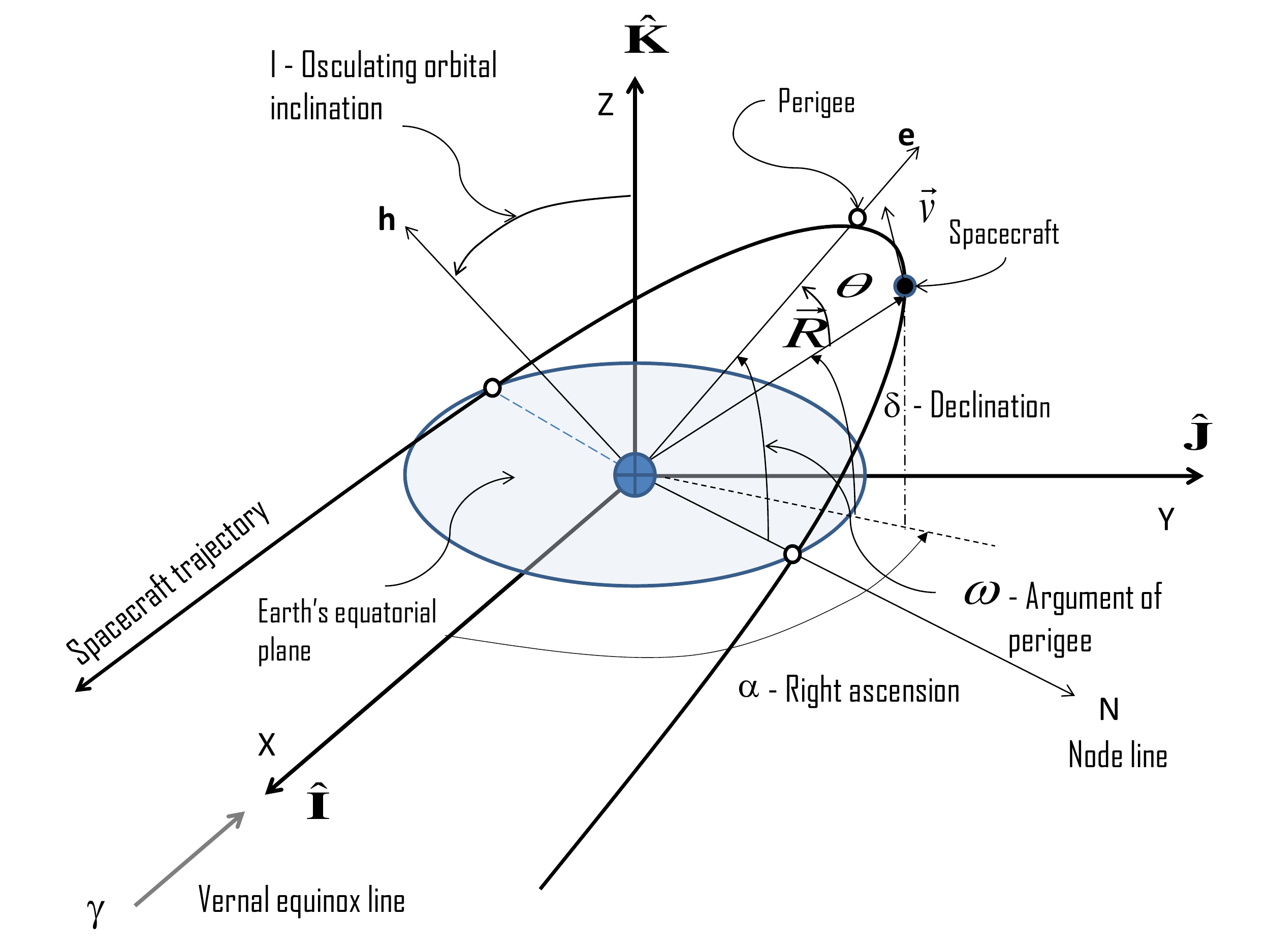}\\
  \caption{Planetary flyby by a spacecraft in the geocentric equatorial frame and the orbital elements. $\mathbf{h}$ is the angular momentum normal to the plane of the orbit and $\mathbf{e}$ is the eccentricity vector. $I$ denotes the osculating orbital inclination to the equator of date, $\omega$ is the osculating argument of the periapsis along the orbit from the equator of date.}\label{Fig2}
\end{figure}

It is implicit into Eq.~\ref{eq7b} the action of the vector potential over a given body, besides the $E$ and $B$-fields, a term analogue to a rotational electric field. Fig.~\ref{Fig2} illustrates the typical planetary flyby by a spacecraft in the geocentric equatorial frame and the orbital elements, where $\mathbf{h}$ is the angular momentum normal to the plane of the orbit and $\mathbf{e}$ is the eccentricity vector pointing along the apse line of the arrival hyperbola.

Let us apply the new governing equation to the planetary flyby of a given spacecraft of mass $m$ nearby a planet of mass $M$, as illustrated in Fig.~\ref{Fig2} (see, e.g., Ref.~\cite{OrbMec}). Hence, in cylindrical geometry, and taking into account the TTC effect alone, Eq.~\ref{eq7b} becomes (Fig.~\ref{Fig2} shows the Earth flyby geometry):
\begin{equation}\label{eq15d}
m \frac{d v_{\theta}}{d t} = m \omega_z A_r \sin I.
\end{equation}
The above Eq.~\ref{eq15d} is written in the geocentric system since it is where the radio tracking data is obtained. Notice that the velocity of the spacecraft relative to the Sun is given by $\mathbf{v}_{sS}=\mathbf{v}_{sP}+ \mathbf{V}_{PS}$, where $\mathbf{v}_{sP}$ is its velocity relative to the planet and $\mathbf{V}_{PS}$ is the velocity of the planet relative to the Sun. But if we consider the term $\mathbf{V}_{PS}$ time-independent, Eq.~\ref{eq15d} gives at the end the azimuthal velocity component of the spacecraft {\it relative} to Earth. Then, if we take due care of the retardation of the gravitational field, it is appropriate to use the gravitational vector potential under the (Li\'{e}nard-Wiechert) form
\begin{equation}\label{eq15dd}
\mathbf{A}(\mathbf{r},t) = \frac{G}{c^2} \frac{M \mathbf{v}_{sP}}{\mid \mathbf{r}-\mathbf{r}' \mid \left(1 - \frac{\mathbf{v}_{sP} \cdot \mathbf{n}'}{c} \right)}.
\end{equation}
Here, $\mathbf{r}$ is the vector position of the planet (e.g., Earth) and $\mathbf{r}'$ is the vector position of the spacecraft, both in the heliocentric system; $\mathbf{n}'$ is the unit vector $(\mathbf{r} -\mathbf{r}')/R$, with $R=\mid \mathbf{r}-\mathbf{r}' \mid$, see Fig.~\ref{Fig2}. We assume that  $\mathbf{V}_{PS}=V_{PS}\mathbf{J}$ and that the planet moves perpendicularly to the vernal line (the Sun is located on the side of the axis $-\mathbf{I}$), along the $\mathbf{J}$ axis (see Fig.~\ref{Fig2}), and therefore $(\mathbf{A} \cdot \mathbf{n}')=A_r$ is the radial component, since what counts in Eq.~\ref{eq15dd} is the relative velocity between spacecraft and planet. The approach velocity vector $\mathbf{v}_{ap}$ is expressed in the approach plane $(\mathbf{i},\mathbf{j},\mathbf{h})$ as follows (the unit vector $\mathbf{i}$ points along the planet direction of motion):
\begin{equation}\label{eq15de}
\mathbf{v}_{ap}=v_{apx} \mathbf{i} + v_{apy} \mathbf{j} + v_{apz} \mathbf{h},
\end{equation}
and the general representation of the spacecraft velocity vector relative to Earth in the direct orthonormal frame is given by
\begin{equation}\label{eq15df}
\begin{array}{ll}
  v_{apx}&=V_P + v_{\infty} \cos (\omega \mp \theta) \\
  v_{apy}&=v_{\infty} \sin(\omega \mp \theta) \\
  v_{apz}&=0.
\end{array}
\end{equation}
Here, $v_{\infty}$ is the excess hyperbolic speed of the spacecraft with respect to the planet. We denote by $\omega_{\oplus}$ the Earth's angular velocity of rotation, $R_{\oplus}$ the Earth's mean radius, $G$ the gravitational constant.
The transit time $dt$ of the spacecraft at the average distance $R_{\oplus}$ (assumed here the radius of the sphere of influence) from the center of the planet (we assume this approximation, since in general the spacecraft altitude is rather smaller than $R_{\oplus}$, see also Ref.~\cite{Anderson 2008}), and we put $d t = d \theta R_{\oplus}/v_{\theta}$, where $v_{\theta}$ is the azimuthal component of the spacecraft velocity, and $d \theta$ denotes the angular deflexion undergone by the spacecraft during the transit time nearby the planet. Expanding Eq.~\ref{eq15dd} to first order in $(\mathbf{v}_{sP} \cdot \mathbf{n}')/c$, we may write Eq.~\ref{eq15d} under the form:
\begin{equation}\label{eq15dg}
dv_{\theta}=\omega_{\oplus} \sin I \frac{GM}{c^2}\frac{V_r}{R_{\oplus}}dt+\omega_{\oplus} \sin I \frac{GM}{c^2}\frac{V_r}{R_{\oplus}}(\mathbf{v}_{sP} \cdot \mathbf{n}')dt,
\end{equation}
or,
\begin{widetext}
\begin{equation}\label{eq15dh}
dv_{\infty} = 2 \omega_{\oplus}R_{\oplus} \sin I \frac{GM}{2R_{\oplus} c^2} d \theta + \frac{2\omega_{\oplus}r_{\oplus}}{c}\frac{GM}{2R_{\oplus}c^2} v_{\infty} \sin(\omega \mp \theta) \sin I d \theta
\end{equation}
\end{widetext}
since $(\mathbf{v}_{sP} \cdot \mathbf{n}')=v_{\infty} \sin (\omega \mp \theta)$ and the radial component of the (relative) velocity is $V_r=\sqrt{v_x^2+(v_y-V_P)^2}=v_{\infty}$. In order to simplify further Eq.~\ref{eq15dh} we may use now the principle of the energy of inertia, which states that the gravitational energy of the spacecraft on the surroundings of the planet must be equal to its energy content according to Einstein formula, e.g., the field itself carries mass, and hence $GM m /2R_{\oplus}=mc^2$. As a result from this equivalence, the velocity variation is independent of the mass of the planet, remaining dependent of its radius, angular velocity and the orbital inclination. Instead to integrate in $\theta$ we may consider the connection between $\theta$ with the declination angle $\delta$ using the trigonometric relationship (see Fig.~\ref{Fig2}):
\begin{equation}\label{eq15fg}
\sin (\omega \mp \theta) \sin I = \sin \delta,
\end{equation}
where $I$ denotes the osculating orbital inclination to the equator of date, $\omega$ is the osculating argument of the perigee along the orbit from the equator of date. This change allows to rewrite Eq.~\ref{eq15dh} in the form of a first-order non-linear differential equation
\begin{equation}\label{eq15di}
\frac{dv_{\infty}}{d \theta}=2 \omega_{\oplus} R_{\oplus} \sin I + K v_{\infty} \sin \delta (\theta),
\end{equation}
where we have now put $v_{\theta}=v_{\infty}$ and noting that $\delta=\delta(\theta)$. It is worth mentioning that the first constant term of Eq.~\ref{eq15di} cancels out when calculating the velocity change $\Delta v_{\infty}$. Therefore, we obtain:
\begin{equation}\label{eq15ff}
\int \frac{d v_{\infty}}{v_{\infty}} = \ln \frac{v_{\infty,f}}{v_{\infty,i}} \approx \frac{\Delta v_{\infty}}{v_{\infty}} = K (\cos \delta_{i} - \cos \delta_{f}).
\end{equation}
Here, $v_{\infty}$ denotes the azimuthal speed of the spacecraft in a position faraway from the planetary influence ($R \to \infty$), $K \equiv 2R_{\oplus} \omega_{\oplus}/c$ is the distance-independent factor, $\delta_i$ and $\delta_f$ denote the initial and final declination angles on the celestial sphere. Eq.~\ref{eq15ff} coincides with the heuristic formula proposed by Anderson~\cite{Anderson_2003}, fitting well for spacecrafts below 2000 km of altitude and has been so far adjusted to high altitudes flyby~\cite{Nieto_2009}.

According to the present analysis the flyby anomaly may have the following causes: i) a drag effect from the planet by means of a Coriolis-like force that push or pull the spacecraft (different from frame dragging, which is debatable~\cite{Allen 2010}); ii) a retarded effect from the gravitational field due to rotation of the planet. The known result, obtained by fitting with experimental data is $\Delta v_{\infty}/v_{\infty} = K(\cos \delta_i - \cos \delta_f)$, where $K=2\omega_{\oplus}R_{\oplus}/c=3.099 \times 10^{-6}$~\cite{Nieto 2009,Diacu 2009,Anderson 2008}.

The dependency on the term $\sin I$ indicates that there is no anomalous acceleration when the inclination angle $I$ is equal to zero. This result is consistent with the data of Table~\ref{tab:table1} collecting the orbital and anomalous dynamical parameters of five Earth flybys as presented in Ref.~\cite{Anderson_2007}. For example, Cassini Earth flyby has no registered data for it just because there is no anomaly; by the contrary, when $I \sim 90\,^{\circ}$, as is the case of NEAR, the variation boost to a higher value $\Delta v_{\infty} =13.46 \pm 0.13$ mm$/$s. Moreover, from the results obtained we see why, due to the vectorial nature of the topological torsion current (and its dependency on the inclination angle), the anomaly can either increase or decrease depending if the spacecraft encounters Earth on the leading or trailing side of its orbital path.

\begin{table*}
\caption{\label{tab:table1} Orbital and anomalous dynamical parameters of five Earth flybys. $b$ is the impact parameter, $A$ is the altitude of the flyby, $I$ is the inclination, $\alpha$ is the right ascension, $\delta$ is the declination of the incoming (i) and outgoing (f) osculating asymptotic velocity vectors. $m$ is the best estimate of the total mass of the spacecraft during the flyby. $v_{\infty}$ is the asymptotic velocity; $\Delta v_{\infty}$ is the increase in the asymptotic velocity of the hyperbolic trajectory. Source: Ref.~\cite{Anderson_2008,Anderson_2007}.}
\begin{ruledtabular}
\begin{tabular}{lllll}
Quantity         &  Galileo (GEGA1)  & NEAR   & Cassini   & Rosetta \\ \hline
b (km$/$s)       &  11,261           & 12,850 & 8,973     & 22,680.49\\
A (km)           &  956.063          & 532.485& 1171.505  & 1954.303\\
I ($\,^{\circ}$)         & 142.9             & 108.0  & 25.4      & 144.9   \\
m (kg)           & 2497.1            & 730.4  & 4612.1    & 2895.2  \\
$\alpha$ ($\,^{\circ}$)  & 163.7             & 240.0  & 223.7     & 269.894 \\
$\delta$ ($\,^{\circ}$)  & 2.975             & -15.37 & -11.16    & -28.185 \\
$\Delta v_{\infty}$ (mm$/$s)& 3.92 $\pm$ 0.08&13.46 $\pm$0.13  & ... & 1.82 $\pm$ 0.05 \\
\end{tabular}
\end{ruledtabular}
\end{table*}

We may conclude from the above that the variational method proposed in Ref.~\cite{Pinheiro_SR} constitutes a powerful alternative approach to tackle problems in the frame of gravito-electromagnetic rotating systems. The emergence of a new force term - the topological torsion current - offers a simple explanation for the flyby anomaly, in fact resulting from a combined slingshot effect (which is not identifiable to frame-dragging) with retardation effects due to the non-instantaneous character of the gravitational force. In addition, the TTC may be well the missing fourth element of force that might be expected on logical and axiomatic point of view, establishing an operational relationship between linear momentum $p$ and angular motion $\theta$. A deeper understanding of the trajectory of the Near-Earth Objects, like asteroids and comets, raising significant concerns, needs a change in the standard assumptions and certainly a deeper understanding of the TTC contribution to the gravitational force will be instrumental when accessing their trajectories.

\vspace*{1cm}

The author gratefully acknowledge financial support from the Portuguese Funda\c{c}\~{a}o para a Ci\^{e}ncia e a Tecnologia (FCT).

\bibliographystyle{apsrev}
\bibliography{Doc2}
\end{document}